\documentclass[twocolumn,english]{revtex4}
\usepackage[T1]{fontenc}
\usepackage[latin9]{inputenc}
\usepackage{float}
\usepackage{amsmath}
\usepackage{amssymb}
\usepackage{graphicx}
\usepackage{esint}

\makeatletter

\providecommand{\tabularnewline}{\\}

\@ifundefined{textcolor}{}
{%
 \definecolor{BLACK}{gray}{0}
 \definecolor{WHITE}{gray}{1}
 \definecolor{RED}{rgb}{1,0,0}
 \definecolor{GREEN}{rgb}{0,1,0}
 \definecolor{BLUE}{rgb}{0,0,1}
 \definecolor{CYAN}{cmyk}{1,0,0,0}
 \definecolor{MAGENTA}{cmyk}{0,1,0,0}
 \definecolor{YELLOW}{cmyk}{0,0,1,0}
}

\newcommand{\be}{\begin{equation}}\newcommand{\ee}{\end{equation}}\newcommand{\ba}{\begin{array}}\newcommand{\ea}{\end{array}}\newcommand{\bea}{\begin{eqnarray}}\newcommand{\eea}{\end{eqnarray}}

\newcommand{\ket}[1]{|#1\rangle}

\newcommand*{\Scale}[2][4]{\scalebox{#1}{$#2$}}%

\numberwithin{lemma}{section}\numberwithin{corol}{section}\numberwithin{prop}{section}\numberwithin{dfn}{section}

\usepackage{babel}

\usepackage{babel}

\usepackage{babel}

\usepackage{babel}

\usepackage{babel}

\usepackage{babel}

\usepackage{babel}

\usepackage{babel}

\usepackage{pgfplots}

\usepackage{babel}

\usepackage{babel}

\makeatother

\usepackage{babel}
\begin{document}

\title{Blackbox Quantization of Superconducting Circuits using exact Impedance
Synthesis}

\author{Firat Solgun\textsuperscript{1,2}, David W. Abraham\textsuperscript{3},
and David P. DiVincenzo\textsuperscript{1,2,4}}

\affiliation{1 Institute for Quantum Information, RWTH Aachen, Germany }

\affiliation{2 J{ü}lich-Aachen Research Alliance (JARA), Fundamentals of Future
Information Technologies, Germany}

\affiliation{3 IBM T.J. Watson Research Center, Yorktown Heights, NY 10598, USA}

\affiliation{4 Peter Gr{ü}nberg Institute: Theoretical Nanoelectronics, Research
Center J{ü}lich, Germany}
\begin{abstract}
We propose a new quantization method for superconducting electronic
circuits involving a Josephson junction device coupled to a linear
microwave environment. The method is based on an exact impedance synthesis
of the microwave environment considered as a blackbox with impedance
function $Z\left(s\right)$. The synthesized circuit captures dissipative
dynamics of the system with resistors coupled to the reactive part
of the circuit in a non-trivial way. We quantize the circuit and compute
relaxation rates following previous formalisms for lumped element
circuit quantization. Up to the errors in the fit our method gives
an exact description of the system and its losses. 
\end{abstract}
\maketitle
\newpage{}

The increase in $Q$-factors of superconducting qubits and cavities
requires highly accurate models for their design, optimization and
predictability. The common approach to model such systems has been
to use Jaynes-Cummings type Hamiltonians borrowed from quantum optics.
However several problems arise like convergence issues when one wants
to include higher levels of superconducting qubits or higher modes
of cavities in such models \cite{Blaisunpub}.

To remedy those issues a method is proposed in \cite{blackbox} to
derive Hamiltonians and compute relaxation rates for superconducting
circuits. In this method the linear electromagnetic environment shunting
the Josephson junction, as extracted, for example, using microwave
simulation software, is lumped together with the junction's linear
inductance, to give a ``blackbox'' impedance function $Z_{sim}\left(\omega\right)$.
This response is then fitted, pole by pole, to an analytic function
$Z\left(\omega\right)$. Then an approximate version of Foster's theorem
\cite{Foster} in the low loss limit \cite{PMC}, applied to $Z\left(\omega\right)$,
gives an equivalent circuit as a series connection of resonant $RLC$
stages, one stage for each term in the partial fraction expansion
of $Z\left(\omega\right)$. In this method, which we refer to as the
``lossy Foster'' method, $Q$ factors for each resonant mode are
computed using $Q_{p}=\frac{\omega_{p}}{2}\frac{Im\left[Y'\left(\omega_{p}\right)\right]}{Re\left[Y\left(\omega_{p}\right)\right]}$
where $\omega_{p}=\left(L_{p}C_{p}\right)^{-1}$ and $Y=Z^{-1}.$
The lifetime of the mode is given by $T_{p}=Q_{p}/\omega_{p}$.

Lossy Foster, while simple to apply, is not always accurate or even
well-conditioned. Terms in the partial-fraction expansion of $Z\left(\omega\right)$
do not always correspond to stages of a physical circuit \cite{Guillemin}.
As Brune showed \cite{Brune}, the property that an impedance function
must have to correspond to a passive physical network is termed ``PR
(Positive-Real)'' this property is an important theme of the present
paper. We note that even if all terms in the expansion of $Z\left(\omega\right)$
are individually $PR$, one might still need to remove terms by inspection
to get a better fit, making the method dependent on ad-hoc decisions.
As applied in \cite{blackbox}, lossy Foster parameters are dependent
not only on the properties of the electromagnetic environment, but
also on the precise value of the junction inductance.

In this paper we propose a new method to derive, from first principles,
the Hamiltonian of a system consisting of a single Josephson junction
connected to a linear microwave environment. As in \cite{blackbox},
we will focus on the example involving a transmon qubit coupled to
a 3D microwave cavity. We also treat the electromagnetic environment
that the junction sees as a black box with an impedance $Z_{sim}$.
To get $Z_{sim}$ we first simulate the cavity system (not including
the linear part of the Josephson inductance) and fit the numerical
impedance response to a rational function $Z\left(s\right)$ 
\begin{equation}
Z\left(s\right)=\frac{n\left(s\right)}{d\left(s\right)}=\underset{k}{\sum}\frac{R_{k}}{s-s_{k}}+d+es\label{eq:partial-fraction-rational}
\end{equation}
(here $s$ is the Laplace variable) using a well established technique
\cite{Vector Fitting}. We then apply the formalism discovered by
Brune \cite{Brune} to synthesize a circuit that has {\em exactly}
the impedance $Z\left(s\right)$ across its terminals. We call the
synthesized circuit the ``Brune circuit''. Since the Brune circuit
has a non-trivial topology, we resort to \cite{BKD,Burkard} to derive
its Hamiltonian and compute relaxation rates. Our method, unlike the
previous lossy Foster approach \cite{blackbox}, involves no approximation
in circuit synthesis. Hence the accuracy of our Hamiltonian and dissipation
analysis give an exact description except for very small errors, introduced
in fitting, which are inevitable in both approaches.

\begin{figure}
\begin{centering}
\includegraphics[scale=0.75]{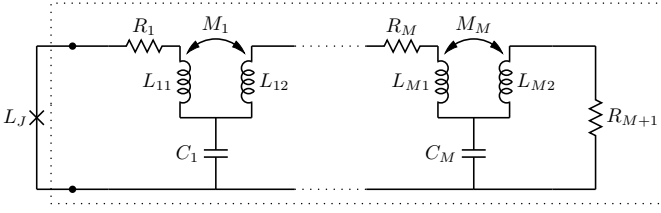} 
\par\end{centering}

\caption{\label{fig:Brune-circuit}Brune circuit (in dotted box) shunted by
a Josephson junction. The analysis of this circuit is extensively
discussed in \cite{SM}, Secs. II and III.}
\end{figure}

After obtaining the rational function fit Eq.~\eqref{eq:partial-fraction-rational}
to $Z\left(s\right)$ (details of which are described below), we use
results from electrical circuit synthesis theory to obtain a lumped
element circuit having exactly this impedance. Brune \cite{Brune}
showed that any impedance response $Z\left(s\right)$ satisfying the
$PR$ conditions can be realised with a finite electric circuit. He
gave an algorithm to find such a lumped element circuit admitting
the $PR$ impedance function $Z\left(s\right)$. This extends Foster's
original work \cite{Foster}, which applies only to lossless networks.
For details of Brune's algorithm see \cite{SM}, Sec. III.B; see also
\cite{Guillemin}. Applying Brune's algorithm to $Z\left(s\right)$
gives a lumped circuit of the form shown in Fig. \ref{fig:Brune-circuit}.

Brune's circuit consists of $M$ stages each containing a tightly-coupled
inductor pair ($M_{j}=\sqrt{L_{j1}L_{j2}}$), a capacitor $C_{j}$,
and a series resistor $R_{j}$. 
As shown below, this interleaving of $M$ lossless stages with $\left(M+1\right)$
resistors results in non-trivial coupling between modes of the circuit
and the dissipative environment represented by these resistors.

We quantize the Brune circuit following the formalism of \cite{Burkard}.
For the Caldeira-Leggett treatment of resistors we refer to \cite{BKD}.
Here we present results of the derivation in \cite{SM}, Sec. II;
we find a Lagrangian $\mathcal{L}_{S}$ (or equivalently a Hamiltonian
$\mathcal{H}_{S}$) corresponding to a 1D chain of interacting oscillator
degrees of freedom: 
\begin{equation}
\mathcal{L}_{S}=\frac{1}{2}\mathbf{\boldsymbol{\dot{\mathbf{\Phi}}}}^{T}\mathbf{\mathcal{C}}\boldsymbol{\dot{\Phi}}-U\left(\boldsymbol{\Phi}\right),\mathcal{H_{S}}=\frac{1}{2}\boldsymbol{Q}^{T}\mathbf{\mathcal{C}^{-1}}\boldsymbol{Q}+U\left(\boldsymbol{\Phi}\right),\label{eq:Hamiltonian}
\end{equation}
where 
\begin{equation}
U\left(\boldsymbol{\Phi}\right)=-\left(\frac{\Phi_{0}}{2\pi}\right)^{2}L_{J}^{-1}cos\left(\varphi_{J}\right)+\frac{1}{2}\boldsymbol{\Phi}^{T}\mathbf{M_{0}}\boldsymbol{\Phi}.\label{eq:Potential-energy-function}
\end{equation}
Here $\mathbf{\Phi}$ is a vector of length $\left(M+1\right)$ whose
entries are linear combinations of branch fluxes in the Brune circuit
(see \cite{SM} Eqs. (18,19) for details). The Josephson phase $\varphi_{J}$
is not an independent coordinate, but it is given in terms of the
variables in the vector ${\bf {\Phi}}$ by $(\Phi_{0}/2\pi)\varphi_{J}=\sum_{j}(-1)^{j+1}\Phi_{j}$.
The chain structure of our representation is evident in the tri-diagonality
of the capacitance and inverse inductance matrices: 
\begin{equation}
\mathcal{C}\mathcal{=}\begin{pmatrix}\Scale[0.8]{C_{1}^{'}} & \Scale[0.8]{t_{1}C_{1}^{'}}\\
\Scale[0.8]{t_{1}C_{1}^{'}} & \Scale[0.8]{t_{1}^{2}C_{1}^{'}+C_{2}^{'}} & \ddots & \text{{\huge0}}\\
 & \ddots & \ddots\\
 & \text{{\huge0}} &  & \Scale[0.8]{t_{M-1}^{2}C_{M-1}^{'}+C_{M}^{'}} & \Scale[0.8]{t_{M}C_{M}^{'}}\\
 &  &  & \Scale[0.8]{t_{M}C_{M}^{'}} & \Scale[0.8]{t_{M}^{2}C_{M}^{'}}
\end{pmatrix},
\end{equation}
\begin{equation}
\mathbf{M_{0}}=\begin{pmatrix}\frac{1}{L'_{1}} & \frac{1}{L'_{1}}\\
\frac{1}{L'_{1}} & \frac{1}{L'_{1}}+\frac{1}{L'_{2}} & \ddots & \text{{\huge0}}\\
 & \ddots & \ddots\\
 & \text{{\huge0}} &  & \frac{1}{L'_{M-1}}+\frac{1}{L'_{M}} & \frac{1}{L'_{M}}\\
 &  &  & \frac{1}{L'_{M}} & \frac{1}{L'_{M}}
\end{pmatrix},
\end{equation}
$C_{j}^{'}=C_{j}/\left(1-t_{j}\right)^{2},$ $L'_{j}=L_{j2}\left(1-t_{j}\right)^{2}$
and $t_{j}=\sqrt{\frac{L_{j1}}{L_{j2}}}$.

Applying Eq. (124) of \cite{BKD} we get the contribution to the relaxation
rate from the resistor $R_{j}$ ($1\leq j\leq M+1$): 
\begin{equation}
\frac{1}{T_{1,j}}=4\left|\left\langle 0\left|{\bf \bar{m}}_{j}\cdot{\bf \Phi}\right|1\right\rangle \right|^{2}J_{j}\left(\omega_{01}\right)\coth\left(\frac{\hbar\omega_{01}}{2k_{B}T}\right)\label{eq:relaxation rate}
\end{equation}
$\ket{0,1}$ are the qubit eigenlevels of the system Hamiltonian Eq.
\eqref{eq:Hamiltonian}. The vector $\bar{{\bf m}}_{j}$ (of length
$(M+1)$) describes the coupling of the system to the environment
representing resistor $R_{j}$; for our Brune circuit this is, for
$1\leq j\leq M$: 
\begin{equation}
\mathbf{\bar{m}}_{j}=\begin{pmatrix}0\\
\vdots\\
0\\
\frac{\left(-1\right)^{j-1}C{}_{j}}{\left(1-t_{j}\right)}\\
\frac{\left(-1\right)^{j}C{}_{j+1}}{\left(1-t_{j+1}\right)}+\frac{\left(-1\right)^{j-1}t_{j}C{}_{j}}{\left(1-t_{j}\right)}\\
\vdots\\
\frac{\left(-1\right)^{M-1}C{}_{M}}{\left(1-t_{M}\right)}+\frac{\left(-1\right)^{M-2}t_{M-1}C{}_{M-1}}{\left(1-t_{M-1}\right)}\\
\frac{\left(-1\right)^{M-1}t_{M}C{}_{M}}{\left(1-t_{M}\right)}
\end{pmatrix}.
\end{equation}
The spectral density corresponding to the bath representing $R_{j}$
is 
\begin{equation}
J_{j}(\omega)={\omega^{3}R_{j}}\left[{1+\omega^{2}R_{j}^{2}\left(\underset{k=j}{\overset{M}{\sum}}C_{k}\right)^{2}}\right]^{-1}.
\end{equation}
For the last resistor $R_{M+1}$, ${\bf \bar{m}}_{M+1}=\left(\begin{array}{cccc}
0 & \cdots & 0 & 1\end{array}\right)^{T}$ and $J_{M+1}\left(\omega\right)={\omega}/{R_{M+1}}$.

\begin{table}
\begin{centering}
\begin{tabular}{|c|c|c|}
\hline 
$k$  & Pole $s_{k}$ $(GHz)$  & Residue $R_{k}$\tabularnewline
\hline 
\hline 
$1$  & $-1.6152\times10^{-6}$  & $8363.13$\tabularnewline
\hline 
$2$,3  & $-0.00110372\pm j6.87473$  & $5.69612\pm j0.00369273$\tabularnewline
\hline 
$4,5$  & $-0.00671733\pm j7.05711$  & $(6.26609\pm j1.34164)\times10^{-5}$\tabularnewline
\hline 
$6,7$  & $-1.34901\pm j8.98453$  & $\left(7.33283\pm j5.61551\right)\times10^{-3}$\tabularnewline
\hline 
$8,9$  & $-0.00272701\pm j12.0048$  & $7.15159\pm j0.0227882$\tabularnewline
\hline 
$10,11$  & $-0.00918635\pm j12.8561$  & $\left(1.98602\pm j0.0134996\right)\times10^{-3}$\tabularnewline
\hline 
$12,13$  & $-1.40214\pm j13.7644$  & $(-8.60807\pm j9.40397)\times10^{-3}$\tabularnewline
\hline 
$14,15$  & $-0.131778\pm j17.7404$  & $23.8075\pm j1.17404$\tabularnewline
\hline 
$16,17$  & $-3.14927\pm88.3524j$  & $\left(1.19527\pm j0.120033\right)\times10^{4}$\tabularnewline
\hline 
\end{tabular}
\par\end{centering}

\caption{\label{tab:Poles-and-residues}Poles and residues for the fit to the
HFSS dataset for $Z_{sim}$ as in the second part of Eq. \eqref{eq:partial-fraction-rational}.}
\end{table}

To show the application of the synthesis method we have just described,
we analyse a dataset produced to analyse a recent 3D transmon experiment
at IBM \cite{3D-IBM}. Our modeling is performed using the finite-element
electromagnetics simulator HFSS\cite{HFSS}. Since the systems we
want to model admit very small loss \cite{3D-transmon,Reagor}, they
are very close to the border which separates stable (passive) systems
from unstable ones. Therefore it is necessary to take care that the
simulation resolution is high enough to ensure the passivity of the
simulated impedance. Otherwise the fitted impedance $Z\left(s\right)$
does not satisfy the $PR$ conditions \cite{Brune} meaning that there
is no passive physical network corresponding to $Z\left(s\right)$.

The physical device that is modelled using HFSS is a rectangular cavity
with a transmon qubit mounted in its center (see \cite{SM}, Figs.
1 and 2). The simulation includes two coaxial ports entering the body
of the cavity symmetrically on either side of the qubit. HFSS is used
to calculate the device's three-port $S$ matrix over a wide frequency
range, from $3.0$ to $15.0$ GHz. The three ports are those defined
by the two coaxial connectors and the qubit terminal pair. That is,
the metal defining the Josephson junction itself is absent from the
simulation, so that its capacitance and (nonlinear) inductance can
be added back later as a discrete element as in Fig. 1. The conversion
from the $S$ matrix to $Z_{sim}$ is calculated using standard formulas
\cite{Newcomb,Pozar}, in which it is assumed that the two coaxial
ports are terminated with a matched ($Z_{0}$=50$\Omega$) resistor.
We have confirmed that the lossy part of the resulting impedance is
mostly determined by these port terminations, rather than by the (physically
rather inaccurate) HFSS model of cavity-metal losses; this is consistent
with the $Q$ of the system being determined by its external couplings
\cite{3D-IBM}.

To obtain the fitted rational impedance function $Z(s)$ as in Eq.
\eqref{eq:partial-fraction-rational}, we use the MATLAB package Vector
Fitting \cite{Vector Fitting}. Vector Fitting is an algorithm to
approximate a sampled impedance/admittance response by a rational
function. It takes a dataset over sampled frequency points, and the
number of poles required for the fit, as its input and gives a set
of poles and residues as its output (See \cite{Zinn} for models with
infinite number of poles). Ref. \cite{Vector Fitting Internals} discusses
details of Vector Fitting. Its passivity enforcement subroutine \cite{VF Passivity Enforcement}
makes sure that the real part of the resulting rational approximation
is positive definite. This feature is crucial for our analysis since
we require the impedance response to be PR (see \cite{SM}, Sec. IIIA)
for the existence of a finite passive network having the same impedance
across its terminals. Note that passivity enforcement may not always
work if the accuracy of the microwave simulation is not high enough
and we have taken care to run the simulation with suitably high resolution.
Applying Vector Fitting to $Z_{sim}$ gives the partial fraction expansion
form in Eq. \eqref{eq:partial-fraction-rational} with the poles $s_{k}$
and residues $R_{k}$ listed in Table \ref{tab:Poles-and-residues},
with $e=0$ and $d=2.80407\Omega$. Note that some of the poles obtained
in the fit have frequencies (imaginary part of $s_{k}$) outside the
range of the simulation data; this is a normal feature of the fitting
routine, used to guarantee a highly accurate fit throughout the entire
simulated frequency band.

We have applied both Brune's algorithm and a lossy Foster analysis
to our fitted $Z\left(s\right)$. Circuit parameters obtained for
the Brune circuit are listed in Table \ref{tab:Parameter-values-for-Brune-Abraham-circuit}.
We see that the series resistor connected directly to the qubit is
quite tiny -- the qubit is nearly lossless. The progressive increase
of the resistance values in further stages of the circuit does not
imply a large contribution of these resistors to loss, as they are
seen by the qubit only through a kind of $LC$ ``filter''. Indeed,
the strong trend towards increasing impedance from stage to stage
in the Brune network (both in the $R$ and $\sqrt{L/C}$ values) means
that the first few stages of the Brune network already give a good
approximation of the cavity response $Z(s)$.

\begin{table}
\begin{centering}
\begin{tabular}{|c|c|c|c|c|}
\hline 
$i$  & $R_{i}\left(\Omega\right)$  & $C_{i}\left(nF\right)$  & $L_{i1}\left(nH\right)$  & $L_{i2}\left(nH\right)$ \tabularnewline
\hline 
\hline 
$1$  & $5.71974\times10^{-5}$  & $1.17020\times10^{-4}$  & $1.32810\times10^{-1}$  & $3.02058\times10^{1}$ \tabularnewline
\hline 
$2$  & $5.53199\times10^{-2}$  & $2.49081\times10^{-6}$  & $8.75272\times10^{1}$  & $3.74225\times10^{3}$ \tabularnewline
\hline 
$3$  & $1.84087\times10^{2}$  & $6.01727\times10^{-8}$  & $4.12954\times10^{3}$  & $1.98121\times10^{4}$ \tabularnewline
\hline 
$4$  & $1.79021\times10^{4}$  & $1.44153\times10^{-9}$  & $4.56024\times10^{4}$  & $2.67489\times10^{5}$ \tabularnewline
\hline 
$5^{*}$  & $6.57108\times10^{5}$  & $2.01906\times10^{-10}$  & $0$  & $0$ \tabularnewline
\hline 
$6$  & $4.90091\times10^{5}$  & $9.69933\times10^{-12}$  & $1.56173\times10^{7}$  & $1.55436\times10^{7}$ \tabularnewline
\hline 
$7$  & $4.14678\times10^{7}$  & $1.64015\times10^{-12}$  & $3.09821\times10^{8}$  & $3.1134\times10^{8}$ \tabularnewline
\hline 
$8$  & $2.33793\times10^{7}$  & $6.32007\times10^{-11}$  & $4.74168\times10^{6}$  & $1.95174\times10^{6}$ \tabularnewline
\hline 
$9$  & $1.22342\times10^{8}$  & $1.70536\times10^{-11}$  & $7.42302\times10^{6}$  & $1.10608\times10^{7}$ \tabularnewline
\hline 
\end{tabular}
\par\end{centering}

\begin{centering}
$R_{10}=6.35712\times10^{8}\Omega$ 
\par\end{centering}

\caption{\label{tab:Parameter-values-for-Brune-Abraham-circuit}Parameter values
for synthesized Brune circuit. Note the strong (orders of magnitude)
increase in impedance (in $R$ and $\sqrt{L/C}$ values) as we go
deep in the circuit. $5^{th}$ stage is degenerate treated in more
detail in \cite{SM}, Sec. II.A.}
\end{table}

\begin{figure}[H]
\begin{centering}
\includegraphics[width=1\columnwidth]{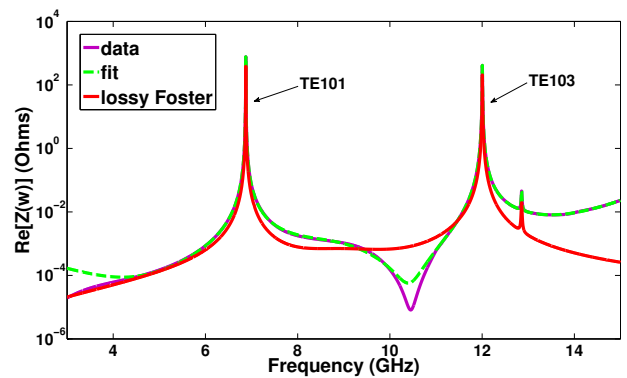} 
\par\end{centering}

\caption{\label{fig:open-circuit-response} Real part of open-circuit response.
Dotted green is open-circuit response for the Brune circuit which
we identify with the open-circuit fit. Solid magenta is the simulated
response. Red is the response of lossy Foster circuit. $TE101$ and
$TE103$ are the resonances associated with classical rectangular
cavity modes\cite{Fields-and-Waves}.}
\end{figure}

In fitting our data with the lossy Foster method (see \cite{SM})
one must be careful about residues with negative real parts or significant
imaginary parts. Note that one cannot apply the lossy Foster approximation
to terms corresponding to poles 12 and 13 in Table \ref{tab:Poles-and-residues}
since they have residues with negative real parts --- there is no
physical network to approximate those terms alone. We also drop DC
and high-frequency terms corresponding to poles 1 and $14-17$ respectively:
such a choice gives a better approximation for the real part of the
impedance in the frequency band of interest. Thus, the best approximating
Foster network consists of five RLC stages, representing the ten remaining
pole pairs.

\begin{figure}
\begin{centering}
\includegraphics[width=1\columnwidth]{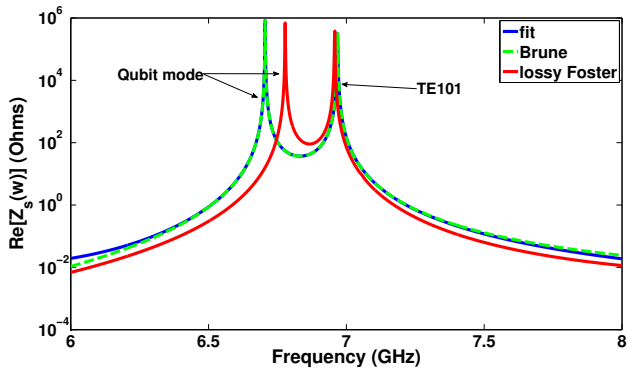} 
\par\end{centering}

\caption{\label{fig:Real-part-of-impedance} Real part of impedance in a small
range of frequencies around the qubit pole ($f_{qb}=6.7052$ GHz where
$f_{qb}$ is the qubit resonance for the exact fit) for the system
shunted (with impedance $Z_{s}$) by a linear inductance $L_{J}=4.5\, nH$
representing the Josephson junction for three different cases. The
$TE101$ mode is not strongly affected by the presence of $L_{J}$.}
\end{figure}

In Fig. \ref{fig:open-circuit-response} we compare these open-circuit
impedances, as represented by the Brune and lossy-Foster methods,
over the full range of our simulation data. The Foster representation
clearly captures the main features of the response, notably the two
classical box resonances of the cavity. But in finer details, especially
far away from the resonances, the Brune representation, which is essentially
indistinguishable from the fit obtained from Vector Fitting, matches
much better than the best lossy Foster circuit.


We now show the improvements that can be expected by using the Brune
circuit when representing the dynamics of the qubit-cavity system.
Here we perform only simple calculations involving a harmonic qubit
(i.e., one represented by a linear inductance $L_{J}$), but our results
give evidence that the Brune circuit will provide high-quality predictions
even for more complex, strongly anharmonic qubits. In Fig. \ref{fig:Real-part-of-impedance}
we show the lossy part of the impedance when the cavity is shunted
by a linear inductance $L_{J}=4.5\, nH$. The fundamental cavity resonance
(TE101) is not significantly changed from the open circuit case, but
the qubit appears as a new pole in the response. This ``qubit pole''
is again very accurately represented by the Brune circuit; however,
using the lossy-Foster circuit derived from the open circuit case
above, the qubit pole is significantly misplaced, by about 100MHz.

\begin{figure}
\begin{centering}
\includegraphics[width=1\columnwidth]{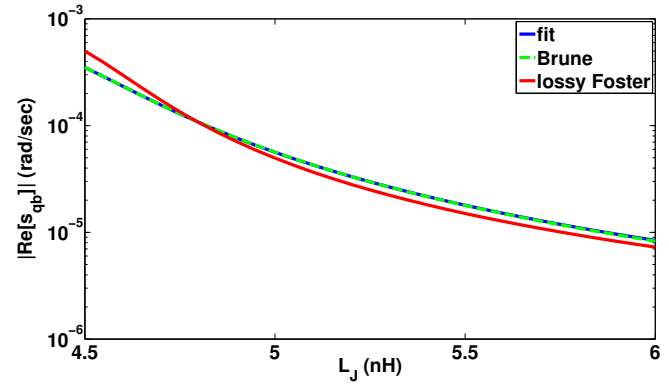} 
\par\end{centering}

\caption{\label{fig:Real-part-of-qubit-pole} Magnitude of the real part of
qubit pole $s_{qb}$ as a function of linear inductance representing
the Josephson junction shunting the system for three different cases:
exact fit for the system shunted by the linear inductance, Brune circuit
shunted by the linear inductance and lossy Foster circuit shunted
by the linear inductance. $T_{1}$ relaxation rate of the qubit is
given by $T_{1}^{-1}=\omega_{qb}/Q_{qb}$, where the quality factor
$Q_{qb}=\omega_{qb}/\left|\xi_{qb}\right|$ with $\xi_{qb}=Re\left[s_{qb}\right]$,
and $\omega_{qb}=Im\left[s_{qb}\right]$ is the frequency of the qubit
mode.}
\end{figure}

Of course, in current applications of the Foster approach \cite{blackbox},
one can do much better by refitting the Foster form with the linear
inductance included in the response, and thus adding a new RLC stage
to explicitly represent the qubit pole. This is an effective strategy,
but the results in Fig.~\ref{fig:Real-part-of-qubit-pole} indicate
its limitations.

Here we compare the use of the Brune and (fixed) lossy-Foster circuit
in giving the real part of the qubit pole, which is proportional to
the relaxation rate $1/T_{1}$ Eq. \eqref{eq:relaxation rate}, as
the inductance $L_{J}$ is varied. We see again that the Brune circuit
matches the ``fit'' result, obtained directly from the HFSS data,
very closely. The deviations of the lossy-Foster result are up to
20\%, and the decrease of the loss rate with $L_{J}$ is significantly
underestimated. This suggest that no single lossy-Foster network,
incorporating some fixed amount of linear inductance, will be able
to match this trend.

Thus, while the Foster approach has been of considerable value in
modelling nearly harmonic qubits like transmons \cite{blackbox},
it appears that the exactness of the Brune approach will be of real
value as we consider other, more anharmonic cavity-coupled qubits.
A clear application in this direction will be the cases of fluxonium
\cite{Pop} or flux qubits \cite{Saclay-Flux-Qubit} -- our approach
should provide a highly accurate multi-mode Hamiltonian for modelling
dynamics in those cases. As we move also to multi-qubit, multi-port
modelling problems, we are hopeful that application of further electrical
theories, developed actively for problems of network synthesis in
the decades after Brune's work, will prove very useful in providing
new modelling techniques for contemporary quantum computer devices.




We thank Gianluigi Catelani for a critical reading of this manuscript.
We are grateful for support from the Alexander von Humboldt foundation.

\begin{widetext}\newpage{}

\begin{center}

{\Large\bf Supplementary Material for "Blackbox Quantization of Superconducting Circuits using exact Impedance Synthesis"}

\bigskip{}

Firat Solgun\textsuperscript{1,2}, David W. Abraham\textsuperscript{3},
and David P. DiVincenzo\textsuperscript{1,2,4}\bigskip{}

1 \textit{Institute for Quantum Information, RWTH Aachen, Germany} 

2 \emph{J{ü}lich-Aachen Research Alliance (JARA), Fundamentals of
Future Information Technologies, Germany}

\emph{3 IBM T.J. Watson Research Center, Yorktown Heights, NY 10598,
USA}

\emph{4 Peter Gr{ü}nberg Institute: Theoretical Nanoelectronics,
Research Center J{ü}lich, Germany}\bigskip{}

\begin{quote}
In these notes we present details of the HFSS simulation, a full derivation
(based on the formalism in \cite{BKD,Burkard}) of the Brune circuit
Hamiltonian and relaxation rate expressions. We also discuss the definition
of PR (Positive-Real) functions, Brune's algorithm and the ``lossy
Foster'' method in detail.
\end{quote}
\end{center}

\section{Device Simulation}

The simulated device is a 3D transmon, inserted with appropriate antenna
structures into the middle of a rectangular superconducting (aluminium)
box cavity, which is standard is several labs presently for high-coherence
qubit experiments. Fig. \ref{fig:cavity-1} shows a perspective rendering
of the device, and Fig. \ref{fig:cavity-2} shows an intensity map
of the fundamental mode of the cavity.

\begin{figure}[H]
\begin{centering}
\includegraphics[scale=0.5]{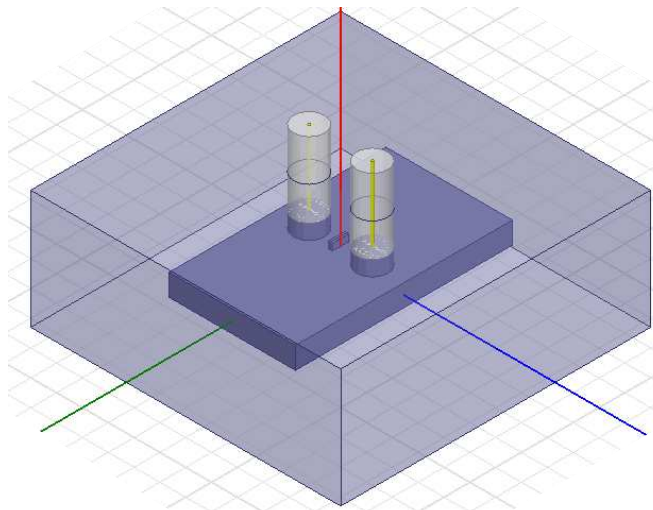} 
\par\end{centering}

\caption{\label{fig:cavity-1}Geometry of the 3D transmon qubit simulated in
HFSS. Light blue is perfect conductor and dark blue is the vacuum.
The qubit port terminals are defined on a dielectric substrate located
at the position of the red line. Two coaxial ports are positioned
symmetrically on each side of the substrate. The cavity dimensions
are $(height,\: length,\: width)=(4.2mm,\:24.5mm,\:42mm)$.}
\end{figure}

\begin{figure}
\begin{centering}
\includegraphics[scale=0.5]{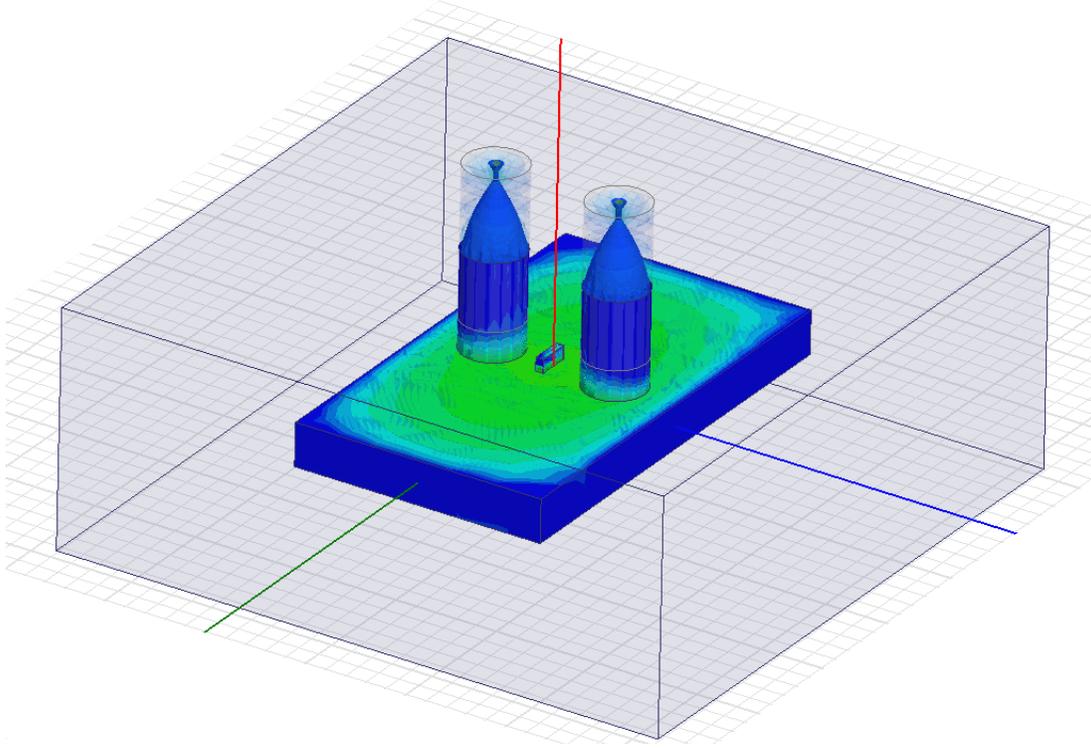} 
\par\end{centering}

\centering{}\caption{\label{fig:cavity-2}Fundamental mode (the $TE101$ mode) of the cavity
with frequency $f_{TE101}=6.875GHz$$ $. Green color indicates electric
field regions of higher magnitude compared to blue regions. }
\end{figure}

\section{Quantization of the Brune circuit}

\begin{figure}
\begin{centering}
\includegraphics{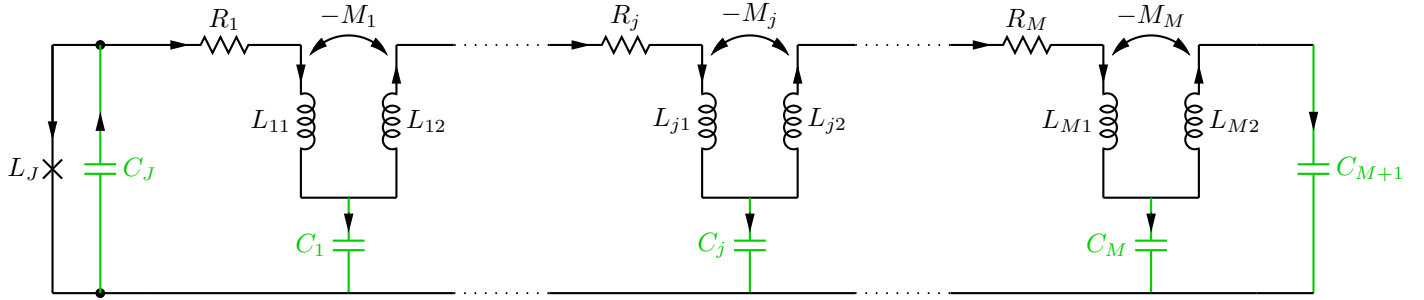} 
\par\end{centering}

\caption{\label{fig:Modified-Brune-circuit}Modified Brune circuit. Tree branches
are shown in black and chord branches are shown in green. Current
directions are chosen to have the matrix $ $$\mathcal{F}_{C}$ in
Eq. \eqref{eq:FC-matrix} with all positive entries.}
\end{figure}

An augmented form of the Brune circuit is shown in Fig. \ref{fig:Modified-Brune-circuit}.
The last resistor $R_{M+1}$ is replaced with a capacitor $C_{M+1}$.
It will be included in our analysis later through the substitution
$C_{M+1}\leftarrow1/(i\omega R_{M+1})$. We will compute its dissipative
effect referring to the equation of motion Eq. (61) in \cite{BKD}.
We also add a formal capacitance $C_{J}$ shunting the Josephson junction.
This is required for a non-singular capacitance matrix if there are
no degenerate stages (see Section \eqref{sub:Brune's-method}). Coupled
inductors in the circuit in Fig. \ref{fig:Modified-Brune-circuit}
satisfy ``tight'' coupling condition $M_{j}=\sqrt{L_{j1}L_{j2}}$.
The inductance matrix $L_{t}$ in Eq. (15) of \cite{Burkard} becomes
singular in the tight coupling limit. To remedy this issue we will
rotate coordinates to eliminate half of degrees of freedom corresponding
to coupled inductor branches. With the ordering $\left(L_{J},L_{12},L_{22},\cdots,L_{M2},L_{11},L_{21},\cdots,L_{M1},R_{1},\cdots,R_{M}\right)$
and $\left(C_{J},C_{1},\cdots,C_{M},C_{M+1}\right)$ for tree and
chord branches respectively (note that right transformer branches
come first and that there are no chord inductors), we construct $\mathcal{F}_{C}$
matrix in Eq. (21) of \cite{Burkard} (To get $\mathcal{F}_{C}$ with
all positive entries we reversed the direction of currents through
and inverted the polarity of voltages across right coupled inductor
branches which requires the update $M_{j}\rightarrow-M_{j}$ for mutual
inductances. See Fig. \ref{fig:Modified-Brune-circuit} for directions
of branch currents and Section \eqref{sub:Brune's-method} for the
definition of the coupled inductor.)

\begin{equation}
\mathcal{F}_{C}=\begin{pmatrix}1 & 1 & 1 & \cdots & 1 & 1\\
 &  & 1 & \cdots & 1 & 1\\
 &  &  & \ddots & \vdots & \vdots\\
 &  & \text{{\huge0}} &  & 1 & 1\\
 &  &  &  &  & 1\\
0 & 1 & 1 & \cdots & 1 & 1\\
 &  & 1 & \cdots & 1 & 1\\
 &  &  & \ddots & \vdots & \vdots\\
 &  & \text{{\huge0}} &  & 1 & 1
\end{pmatrix}\label{eq:FC-matrix}
\end{equation}
where $\mathcal{F}_{C}$ is a ($2M+1)\times(M+2)$ matrix. We then
compute the capacitance matrix in Eq. (22) of \cite{Burkard} as

\begin{equation}
\mathcal{C}_{0}=\mathcal{F}_{C}C\mathcal{F}_{C}^{t}
\end{equation}
where $C$ is the diagonal matrix with capacitances $\left(C_{J},C_{1},\cdots,C_{M},C_{M+1}\right)$
in the diagonal. With the directions chosen for coupled inductor currents
$L_{t}^{-1}$ in Eq. (16) of \cite{Burkard} is written as

\begin{equation}
L_{t}^{-1}=\frac{1}{L_{0}^{2}}\begin{pmatrix}L_{11} &  & \text{{\large0}} & M_{1} &  & \text{{\large0}}\\
 & \ddots &  &  & \ddots\\
\text{{\large0}} &  & L_{1M} & \text{{\large0}} &  & M_{M}\\
M_{1} &  & \text{{\large0}} & L_{12} &  & \text{{\large0}}\\
 & \ddots &  &  & \ddots\\
\text{{\large0}} &  & M_{M} & \text{{\large0}} &  & L_{M2}
\end{pmatrix}
\end{equation}
where $M_{j}=\sqrt{L_{j1}L_{j2}-L_{0}^{2}}$ with $L_{0}>0$ being
a small parameter giving the deviation from the tight coupling limit.
We have

\begin{equation}
\mathcal{G}=\begin{pmatrix}0\\
1_{2M\times2M}
\end{pmatrix}
\end{equation}
and

\begin{align}
M_{0} & =\mathcal{G}L_{t}^{-1}\mathcal{G}^{t}\\
 & =\begin{pmatrix}0 & 0\\
0 & L_{t}^{-1}
\end{pmatrix}
\end{align}

We construct a rotation matrix $U$

\begin{equation}
U=\begin{pmatrix}1 & 0 &  & \cdots &  &  & 0\\
0 & \frac{1}{\sqrt{1+t_{1}^{2}}} &  & \text{{\large0}} & \frac{t_{1}}{\sqrt{1+t_{1}^{2}}} &  & \text{{\large0}}\\
 &  & \ddots &  &  & \ddots\\
\vdots & \text{{\large0}} &  & \frac{1}{\sqrt{1+t_{M}^{2}}} & \text{{\large0}} &  & \frac{t_{M}}{\sqrt{1+t_{M}^{2}}}\\
 & -\frac{t_{1}}{\sqrt{1+t_{1}^{2}}} &  & \text{{\large0}} & \frac{1}{\sqrt{1+t_{1}^{2}}} &  & \text{{\large0}}\\
 &  & \ddots &  &  & \ddots\\
0 & \text{{\large0}} &  & -\frac{t_{M}}{\sqrt{1+t_{M}^{2}}} & \text{{\large0}} &  & \frac{1}{\sqrt{1+t_{M}^{2}}}
\end{pmatrix}
\end{equation}
where $t_{j}=\sqrt{\frac{L_{j1}}{L_{j2}}}$. We now compute $U^{t}M_{0}U$
and truncate it to its upper-left $\left(M+1\right)\times\left(M+1\right)$
sector (by taking $L_{0}\rightarrow0$ limit) which corresponds to
the eigenspace with finite(non-infinite) eigenvalues. After truncation
we get

\begin{equation}
M_{0}^{'}=\begin{pmatrix}0 &  & \text{{\large0}}\\
 & 1/L_{1}\\
 &  & \ddots\\
 & \text{{\large0}} &  & 1/L_{M}
\end{pmatrix}
\end{equation}
where $L_{j}=L_{j1}+L_{j2}$. After transforming $\mathcal{C}_{0}$
by computing $U^{t}\mathcal{C}_{0}U$ and truncating we get $\mathcal{C}_{0}^{'}$.
The matrix $\mathcal{C}_{0}^{'}$ is in general non-zero in all its
entries but below we construct a second transformation matrix $T$
to make both $\mathcal{C}_{0}^{'}$ and $M'_{0}$ band-diagonal

\begin{equation}
T=\begin{pmatrix}1\\
-\frac{\sqrt{1+t_{1}^{2}}}{1-t_{1}} & -\frac{\sqrt{1+t_{1}^{2}}}{1-t_{1}} &  & \text{{\huge0}}\\
 & \frac{\sqrt{1+t_{2}^{2}}}{1-t_{2}} & \frac{\sqrt{1+t_{2}^{2}}}{1-t_{2}}\\
 &  & \ddots & \ddots\\
 & \text{{\huge0}} &  & \left(-1\right)^{M}\frac{\sqrt{1+t_{M}^{2}}}{1-t_{M}} & \left(-1\right)^{M}\frac{\sqrt{1+t_{M}^{2}}}{1-t_{M}}
\end{pmatrix}
\end{equation}
Applying $T$ to $\mathcal{C}_{0}^{'}$ and $M_{0}^{'}$ we get

\begin{align}
\mathcal{C} & =T^{t}\mathcal{C}_{0}^{'}T\label{eq:capacitance}\\
 & \mathcal{=}\begin{pmatrix}C_{J}+C_{1}^{'} & t_{1}C_{1}^{'}\\
t_{1}C_{1}^{'} & t_{1}^{2}C_{1}^{'}+C_{2}^{'} & \ddots & \text{{\huge0}}\\
 & \ddots & \ddots\\
 & \text{{\huge0}} &  & t_{M-1}^{2}C_{M-1}^{'}+C_{M}^{'} & t_{M}C_{M}^{'}\\
 &  &  & t_{M}C_{M}^{'} & t_{M}^{2}C_{M}^{'}+C_{M+1}^{'}
\end{pmatrix}
\end{align}

\begin{align}
\mathbf{M_{0}} & =T^{t}M_{0}^{'}T\\
 & =\begin{pmatrix}\frac{1}{L'_{1}} & \frac{1}{L'_{1}}\\
\frac{1}{L'_{1}} & \frac{1}{L'_{1}}+\frac{1}{L'_{2}} & \frac{1}{L'_{2}} &  & \text{{\huge0}}\\
 & \frac{1}{L'_{2}} & \frac{1}{L'_{2}}+\frac{1}{L'_{3}} & \ddots\\
 &  & \ddots & \ddots\\
 & \text{{\huge0}} &  &  & \frac{1}{L'_{M-1}}+\frac{1}{L'_{M}} & \frac{1}{L'_{M}}\\
 &  &  &  & \frac{1}{L'_{M}} & \frac{1}{L'_{M}}
\end{pmatrix}\label{eq:M0}
\end{align}
where $C_{j}^{'}=C_{j}/\left(1-t_{j}\right)^{2}$, $L'_{j}=L_{j2}\left(1-t_{j}\right)^{2}$.

A Lagrangian $\mathcal{L}_{0}$ (and equivalently a Hamiltonian $\mathcal{H_{S}}$)
can be written as

\begin{equation}
\mathcal{L}_{0}=\frac{1}{2}\mathbf{\boldsymbol{\dot{\mathbf{\Phi}}}}^{T}\mathcal{C}\boldsymbol{\dot{\Phi}}-U\left(\boldsymbol{\Phi}\right),\;\mathcal{H_{S}}=\frac{1}{2}\boldsymbol{Q}^{T}\mathcal{C}^{-1}\boldsymbol{Q}+U\left(\boldsymbol{\Phi}\right)\label{eq:Lagrangian}
\end{equation}
where

\begin{equation}
U\left(\boldsymbol{\Phi}\right)=-\left(\frac{\Phi_{0}}{2\pi}\right)^{2}L_{J}^{-1}cos\left(\varphi_{J}\right)+\frac{1}{2}\boldsymbol{\Phi}^{T}\mathbf{M_{0}}\boldsymbol{\Phi}\label{eq:Potential-energy-function-1}
\end{equation}
$\mathbf{\Phi}$ is the vector of transformed(and truncated) coordinates
of length $\left(M+1\right)$. $ $$\varphi_{L}$ is the phase across
the Josephson junction. One can relate $\mathbf{\Phi}$ to the original
branch fluxes in the Brune circuit by introducing an auxiliary vector
$\mathbf{\Phi'}$ of length $\left(M+1\right)$ and keeping track
of two coordinate transformations $U$ and $T$ applied as follows

\begin{equation}
\mathbf{\Phi}=T^{t}\mathbf{\Phi}'
\end{equation}
with

\begin{align}
\mathbf{\Phi}' & =\left(\Phi_{J},\Phi_{1}',\cdots,\Phi_{M}'\right)\\
 & =U^{t}\left(\Phi_{J},\boldsymbol{\Phi}_{L}\right)^{t}
\end{align}
where

\begin{equation}
\left(\Phi_{J},\boldsymbol{\Phi}_{L}\right)=\left(\Phi_{J},\Phi_{12},\Phi_{22},\cdots,\Phi_{M2},\Phi_{11},\Phi_{21},\cdots,\Phi_{M1}\right)
\end{equation}
is the vector of fluxes of tree branches in the Brune circuit in Fig.
\ref{fig:Modified-Brune-circuit} , $\Phi_{J}=\left(\frac{\Phi_{0}}{2\pi}\right)\varphi_{J}$
and $\Phi_{j}'=\frac{1}{\sqrt{1+t_{j}^{2}}}\left(\Phi_{j2}-t_{j}\Phi_{j1}\right)$
, for $1\leq j\leq M$. Here we assume that the vector $U^{t}\left(\Phi_{J},\boldsymbol{\Phi}_{L}\right)^{t}$
is truncated to its first $ $$\left(M+1\right)$ rows before assignment
to $\mathbf{\Phi}'$. As shown in Fig. \ref{fig:circuit-eigenmodes}
the mode $\Phi_{j}$ of the circuit is a linear combination of four
fluxes across inductors in stage $j$ and $j+1$. More specifically
we can write the $j^{th}$ component of $\mathbf{\Phi}$ for $2\leq j\leq M$
by

\begin{align}
\Phi_{j} & =\left(-1\right)^{j-1}\frac{\sqrt{1+t_{j-1}^{2}}}{1-t_{j-1}}\Phi'_{j-1}+\left(-1\right)^{j}\frac{\sqrt{1+t_{j}^{2}}}{1-t_{j}}\Phi'_{j}\\
 & =\frac{\left(-1\right)^{j-1}}{1-t_{j-1}}\left(\Phi_{j-1,2}-t_{j-1}\Phi_{j-1,1}\right)+\frac{\left(-1\right)^{j}}{1-t_{j}}\left(\Phi_{j2}-t_{j}\Phi_{j1}\right)\label{eq:eigenmode}
\end{align}
For $j=1,\, M+1$ we have $\Phi_{1}=\Phi_{J}-\frac{\sqrt{1+t_{1}^{2}}}{1-t_{1}}\Phi_{1}'=\Phi_{J}-\left(\Phi_{12}-t_{1}\Phi_{11}\right)/\left(1-t_{1}\right)$
and $\Phi_{M+1}=\left(-1\right)^{M}\frac{\sqrt{1+t_{M}^{2}}}{1-t_{M}}\Phi'_{M}=\left(-1\right)^{M}\left(\Phi_{M2}-t_{M}\Phi_{M1}\right)/\left(1-t_{M}\right)$
, respectively. Note that the Josephson phase $\Phi_{J}$ is given
by $\Phi_{J}=\underset{j}{\sum}\left(-1\right)^{j+1}\Phi_{j}$.

\begin{figure}
\begin{centering}
\includegraphics[scale=0.9]{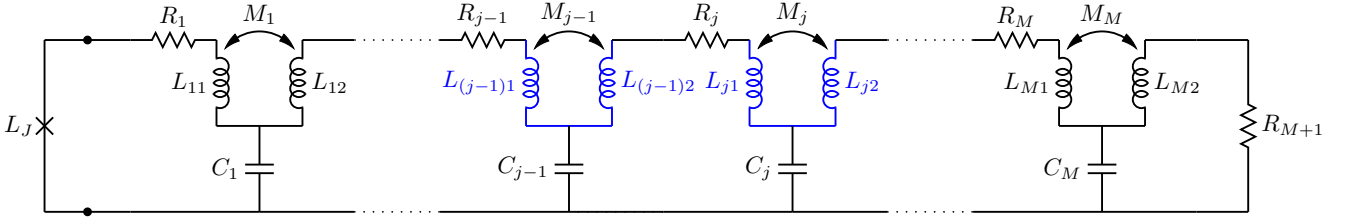} 
\par\end{centering}

\caption{\label{fig:circuit-eigenmodes}The $j^{th}$ mode $\Phi_{j}$ in Eq.
\eqref{eq:Lagrangian} of the Brune circuit. $\Phi_{j}$ is a linear
combination of four branch fluxes $\Phi_{j-1,1},\Phi_{j-1,2},\Phi_{j1},\Phi_{j2}$
across inductors over two consecutive stages, as given by Eq. \eqref{eq:eigenmode}.}
\end{figure}

To treat resistors in Caldeira-Leggett formalism we will first compute
the dissipation matrix $\mathcal{C}_{Z}\left(\omega\right)$ in Eq.
(26) of \cite{Burkard}. We will then interpret the equation of motion
$\left(\mathcal{C}+\mathcal{C}_{Z}\right)*\ddot{\Phi}=-\frac{\partial U}{\partial\Phi}$
in Eq. (29) of \cite{Burkard} as an equation of motion Eq. (61) of
\cite{BKD} by taking the dissipative term to the right-hand side
and writing (in frequency domain) $\mathcal{C}\ddot{\Phi}=-\frac{\partial U}{\partial\Phi}-\omega^{2}\mathcal{C}_{Z}\Phi$.
One can then relate $\mathbf{M}_{d}\left(\omega\right)=\omega^{2}\mathcal{C}_{Z}$
and $K\left(\omega\right)=\omega^{2}\bar{\mathbf{C}}_{Z}\left(\omega\right)$
where $\mathbf{M}_{d}$ and $K\left(\omega\right)$ are given in Eqs.
(72-75) of \cite{BKD}. Then coupling vectors $\mathbf{\bar{m}}$
are identical in both formalisms.

We treat each resistor separately. Applying Eq. (124) of \cite{BKD}
we get the contribution to the relaxation rate from the resistor $R_{j}$
($1\leq j\leq M+1$):

\begin{equation}
\frac{1}{T_{1,j}}=4\left|\left\langle 0\left|\mathbf{\bar{m}_{j}}\cdot\mathbf{\Phi}\right|1\right\rangle \right|^{2}J_{j}\left(\omega_{01}\right)\coth\left(\frac{\hbar\omega_{01}}{2k_{B}T}\right)\label{eq:relaxation-rate}
\end{equation}
$\ket{0,1}$ are the qubit eigenlevels of the system Hamiltonian Eq.
\eqref{eq:Lagrangian}. The vector ${\bf \bar{m}}_{j}$ (of length
$(M+1)$) describes the coupling of the system to the environment
representing resistor $R_{j}$. Note that our use of the non-normalized
coupling vector $\mathbf{\bar{m}_{j}}$ and the flux vector $\mathbf{\Phi}$
implies removal of the factor $\mu\left(\frac{\Phi_{0}}{2\pi}\right)^{2}$
from the definition of the spectral function of the bath $J$ in Eq.
(93) of \cite{BKD} (See Eqs. \eqref{eq:spectral-function} and \eqref{eq:spectral-function-last-resistor}
below).

For $1\leq j\leq M$, using Eqs. (26-28) in \cite{Burkard} we compute

\begin{equation}
\mathbf{\bar{m}_{j}}=\begin{pmatrix}0\\
\vdots\\
0\\
\frac{\left(-1\right)^{j-1}C{}_{j}}{\left(1-t_{j}\right)}\\
\frac{\left(-1\right)^{j}C{}_{j+1}}{\left(1-t_{j+1}\right)}+\frac{\left(-1\right)^{j-1}t_{j}C{}_{j}}{\left(1-t_{j}\right)}\\
\vdots\\
\frac{\left(-1\right)^{M-1}C{}_{M}}{\left(1-t_{M}\right)}+\frac{\left(-1\right)^{M-2}t_{M-1}C{}_{M-1}}{\left(1-t_{M-1}\right)}\\
\frac{\left(-1\right)^{M-1}t_{M}C{}_{M}}{\left(1-t_{M}\right)}
\end{pmatrix}
\end{equation}
where $\mathbf{\bar{m}_{j}}$ are vectors of length $(M+1)$ and

\begin{equation}
\bar{\mathbf{C}}_{Z,j}\left(\omega\right)=-\frac{i\omega R_{j}}{1+i\omega R_{j}\left(\underset{k=j}{\overset{M}{\sum}}C_{k}\right)}
\end{equation}
We then have

\begin{align}
K_{j}\left(\omega\right) & =\omega^{2}\bar{\mathbf{C}}_{Z,j}\left(\omega\right)\\
 & =\frac{i\omega^{3}R_{j}}{1+i\omega R_{j}\left(\underset{k=j}{\overset{M}{\sum}}C_{k}\right)}
\end{align}
Hence

\begin{align}
J_{j} & =Im\left[K_{j}\left(\omega\right)\right]\label{eq:spectral-function}\\
 & =\frac{\omega^{3}R_{j}}{1+\omega^{2}R_{j}^{2}\left(\underset{k=j}{\overset{M}{\sum}}C_{k}\right)^{2}}
\end{align}

To treat last resistor $R_{M+1}$ we first replace $C_{M+1}$ in the
last row of capacitance matrix by $1/(i\omega R_{M+1})$. This gives
a term $-\frac{1}{R_{M+1}}\dot{\varphi}_{M}$ on the right hand side
of the equation of motion in Eq. (29) of \cite{Burkard}. This term
can be treated with \cite{BKD}. It gives rise to a dissipation matrix
$\mathbf{M}_{d}=K_{M+1}\left(\omega\right)\mathbf{\bar{m}}_{M+1}\mathbf{\bar{m}}_{M+1}^{T}$
where $K_{M+1}\left(\omega\right)=\frac{i\omega}{R_{M+1}}$ and $\mathbf{\bar{m}}_{M+1}=\begin{pmatrix}0\\
\vdots\\
0\\
1
\end{pmatrix}$ is a vector with $\left(M+1\right)$ rows. We then have

\begin{equation}
J_{M+1}\left(\omega\right)=Im\left[K_{M+1}\left(\omega\right)\right]=\frac{\omega}{R_{M+1}}\label{eq:spectral-function-last-resistor}
\end{equation}

\subsection{Degenerate case}

As discussed in Appendix \eqref{sub:Brune's-method} Brune's algorithm
may produce degenerate stages. In this text we will only consider
the capacitive degenerate case. Such a case has appeared in the example
circuit we studied as listed in Table II of the main text. We consider
a degenerate case appearing at $k^{th}$ stage. As noted in Section
\eqref{sub:Brune's-method} such a stage corresponds to the limit
of $L'_{k}\rightarrow0$ and $t_{k}\rightarrow0$ . To remove the
singularity we define a transformation

\begin{equation}
T_{d}=\begin{pmatrix}1\\
 & \ddots\\
 &  & 1\\
row & \left(k+1\right)\rightarrow & -1 & -1\\
 &  &  &  & \ddots\\
 &  &  &  &  & -1
\end{pmatrix}
\end{equation}
Applying this tranformation to the matrices $\mathbf{M_{0}}$ and
$\mathcal{C}$ and removing the coordinate of the degenerate stage(this
corresponds to the removal of $\left(k+1\right)^{th}$ row and $\left(k+1\right)^{th}$
column from both matrices) we get

\begin{equation}
T_{d}^{t}\mathbf{M_{0}}T_{d}=\begin{pmatrix}\frac{1}{L'_{1}} & \frac{1}{L'_{1}}\\
\frac{1}{L'_{1}} & \frac{1}{L'_{1}}+\frac{1}{L'_{2}} & \frac{1}{L'_{2}}\\
 & \frac{1}{L'_{2}} & \frac{1}{L'_{2}}+\frac{1}{L'_{3}} & \ddots &  &  & \text{{\huge0}}\\
 &  & \ddots & \ddots\\
 &  &  &  & \frac{1}{L'_{k-1}}+\frac{1}{L'_{k+1}} & \frac{1}{L'_{k+1}}\\
 &  &  &  & \frac{1}{L'_{k+1}} & \frac{1}{L'_{k+1}}+\frac{1}{L'_{k+2}} & \ddots\\
 &  & \text{{\huge0}} &  &  & \ddots & \ddots\\
 &  &  &  &  &  &  & \frac{1}{L'_{M-1}}+\frac{1}{L'_{M}} & \frac{1}{L'_{M}}\\
 &  &  &  &  &  &  & \frac{1}{L'_{M}} & \frac{1}{L'_{M}}
\end{pmatrix}
\end{equation}

$T_{d}^{t}\mathcal{C}T_{d}=$

\begin{equation}
\begin{pmatrix}C_{J}+C_{1}^{'} & t_{1}C_{1}^{'}\\
t_{1}C_{1}^{'} & t_{1}^{2}C_{1}^{'}+C_{2}^{'} & \ddots &  &  & \text{{\huge0}}\\
 & \ddots & \ddots\\
 &  &  & t_{k-1}^{2}C'_{k-1}+\left(C'_{k+1}+C'_{k}\right) & t_{k+1}C'_{k+1}\\
 &  &  & t_{k+1}C'_{k+1} & t_{k+1}^{2}C'_{k+1}+C'_{k+2} & \ddots\\
 &  &  &  & \ddots & \ddots\\
 &  & \text{{\huge0}} &  &  &  & t_{M-1}^{2}C_{M-1}^{'}+C_{M}^{'} & t_{M}C_{M}^{'}\\
 &  &  &  &  &  & t_{M}C_{M}^{'} & t_{M}^{2}C_{M}^{'}+C{}_{M+1}
\end{pmatrix}\label{eq:degenerate-c-matrix}
\end{equation}
Note that the matrices above are of size $M\times M$.

One needs to update also $\bar{\mathbf{m}}$ vectors. To do this we
have to apply the transformation $T_{d}$ to $\bar{\mathbf{m}}$ vectors
and remove the entry corresponding to the degenerate coordinate (i.e.
the $\left(k+1\right)^{th}$ row). Now we define some auxiliary vectors

\begin{equation}
\mathbf{\bar{m}}_{a}\left(j\right)=\begin{pmatrix}0\\
\vdots\\
0\\
j^{th}\; row\longrightarrow\left(-1\right)^{j-1}\frac{C{}_{j}}{\left(1-t_{j}\right)}\\
\left(-1\right)^{j}\frac{C{}_{j+1}}{\left(1-t_{j+1}\right)}+\left(-1\right)^{j-1}t_{j}\frac{C{}_{j}}{\left(1-t_{j}\right)}\\
\vdots\\
\left(-1\right)^{k-2}\frac{C{}_{k-1}}{\left(1-t_{k-1}\right)}+\left(-1\right)^{k-3}t_{k-2}\frac{C{}_{k-2}}{\left(1-t_{k-2}\right)}\\
k^{th}\; row\longrightarrow\left(-1\right)^{k-2}t_{k-1}\frac{C{}_{k-1}}{\left(1-t_{k-1}\right)}\\
0\\
\vdots\\
0
\end{pmatrix}\label{eq:coupling-vector-ma}
\end{equation}

\begin{equation}
\mathbf{\bar{m}}_{b}\left(j\right)=\begin{pmatrix}0\\
\vdots\\
0\\
\left(j-1\right)^{th}\; row\longrightarrow\left(-1\right)^{j}\frac{C{}_{j}}{\left(1-t_{j}\right)}\\
\left(-1\right)^{j+1}\frac{C{}_{j+1}}{\left(1-t_{j+1}\right)}+\left(-1\right)^{j}t_{j}\frac{C{}_{j}}{\left(1-t_{j}\right)}\\
\vdots\\
\left(-1\right)^{M}\frac{C{}_{M}}{\left(1-t_{M}\right)}+\left(-1\right)^{M-1}t_{M-1}\frac{C{}_{M-1}}{\left(1-t_{M-1}\right)}\\
\left(-1\right)^{M}t_{M}\frac{C{}_{M}}{\left(1-t_{M}\right)}
\end{pmatrix}\label{eq:coupling-vector-mb}
\end{equation}

\begin{equation}
\bar{\mathbf{m}}_{C_{k}}=\begin{pmatrix}0 & \cdots & 0 & C_{k} & 0 & \cdots & 0\end{pmatrix}^{t}\label{eq:coupling-vector-mck}
\end{equation}
where $C_{k}$ is in $k^{th}$ row. Now we can write coupling vector
$\bar{\mathbf{m}}_{j}$ to the bath of the resistor $R_{j}$ as a
function of the vectors defined in Eqs. \eqref{eq:coupling-vector-ma},
\eqref{eq:coupling-vector-mb}, \eqref{eq:coupling-vector-mck} above
as

\begin{equation}
\mathbf{\bar{m}}_{j}=\begin{cases}
\bar{\mathbf{m}}_{a}\left(j\right)+\bar{\mathbf{m}}_{C_{k}}+\mathbf{\bar{m}}_{b}\left(k\right) & for\; j<k\\
\bar{\mathbf{m}}_{C_{k}}+\mathbf{\bar{m}}_{b}\left(k\right) & for\; j=k\\
\mathbf{\bar{m}}_{b}\left(j\right) & for\; j>k
\end{cases}
\end{equation}

Note that vectors above are all of length $M$. Spectral densities
$J_{i}\left(\omega\right)$ are the same as in the non-degenerate
case (Eqs. \eqref{eq:spectral-function},\eqref{eq:spectral-function-last-resistor})
for all resistors. Note also that dissipation treatment for the last
resistor $R_{M+1}$ is unaffected since $C{}_{M+1}$ is untouched
in Eq. \eqref{eq:degenerate-c-matrix}.

\section{Brune's method\label{sub:Brune's-method}}

Brune extended\cite{Brune} Foster's\cite{Foster} work to lossy networks.
He formulated necessary and sufficient conditions for a rational function
$Z\left(s\right)$ to correspond to a passive lumped element circuit
including possibly resistors. He coined the term ``positive real
(PR)'' for such functions. He also devised an algorithm to synthesize
a circuit given a PR function $Z\left(s\right)$. Below we define
PR property and describe Brune's algorithm. For more details see \cite{Guillemin}.
In the following we stick with the electrical engineering convention
for the imaginary unit $j=-i$.

\subsection{PR property}

A scalar impedance function $Z\left(s\right)$ is PR if the following
two conditions are met

1) $Z\left(s\right)$ is a rational function which is real for real
values of $s$.

2) $Re\left[Z\left(s\right)\right]\geq0$ for $Re\left[s\right]\geq0$.

The second condition is equivalent to the following

1) No poles lie in the right half plane.

2) Poles on the j-axis have finite positive real residues and are
simple.

3) $Re\left[Z\left(j\omega\right)\right]\geq0$.

\subsection{Brune's algorithm}
\begin{enumerate}
\item If $Z\left(s\right)$ or $Y\left(s\right)=1/Z\left(s\right)$ has
j-axis poles, remove them by realizing terms corresponding to those
poles in the partial fraction expansion. Those terms correspond to
parallel $LC$ resonators(connected in series) in case of $Z\left(s\right)$
poles and series $LC$ resonators(connected in parallel) for $Y\left(s\right)$
poles. Repeat until no j-axis pole is left. 
\item Find $\omega_{1}$ and $R_{1}$ such that $R_{1}=\underset{\omega}{\min}Z\left(j\omega\right)$
and $Z\left(j\omega_{1}\right)=R_{1}$ . Define $Z_{1}\left(s\right)=Z\left(s\right)-R_{1}$
. This step corresponds to the removal of $R_{1}$ in Fig. \ref{fig:Brune-circuit-extraction}. 
\item Define $L_{1}=Z_{1}\left(j\omega_{1}\right)/\left(j\omega_{1}\right)$.
If we extract the inductance $L_{1}$ as shown in Fig. \ref{fig:Brune-circuit-extraction},
$1/\left(Z_{1}\left(s\right)-L_{1}s\right)$ is the admittance corresponding
to the rest of the circuit and has a pole at $s=j\omega_{1}$, hence
we can write 
\begin{equation}
\frac{1}{Z_{1}\left(s\right)-L_{1}s}=\frac{\left(1/L_{2}\right)s}{s^{2}+\omega_{1}^{2}}+\frac{1}{W\left(s\right)}\label{eq:Brune-circuit-extraction-step-3}
\end{equation}

\item The first term in Eq. \eqref{eq:Brune-circuit-extraction-step-3}
corresponding to the pole at $s=j\omega_{1}$ is realized with a shunt
$LC$ branch consisting of inductance $L_{2}$ connected in series
with capacitance $C_{2}=1/\left(L_{2}\omega_{1}^{2}\right)$ as shown
in Fig. \ref{fig:Brune-circuit-extraction}. 
\item $W\left(s\right)$ has a pole at infinity such that 
\end{enumerate}
\begin{equation}
\underset{s\rightarrow\infty}{\lim}W\left(s\right)=-\frac{L_{1}L_{2}s}{L_{1}+L_{2}}=L_{3}s
\end{equation}

This pole is removed by constructing $Z_{2}\left(s\right)=W\left(s\right)-L_{3}s$
which corresponds to connecting in series an inductance of value $L_{3}=-L_{1}L_{2}/\left(L_{1}+L_{2}\right)$.
$Z_{2}\left(s\right)$ is PR with no j-axis poles or zeros and whole
process(steps 1 to 5) can now be applied to $Z_{2}$.

Steps 1 to 5 reduce degrees of both numerator and denominator of $Z\left(s\right)$
by 2 so that the algorithm terminates once a constant $Z_{2}\left(s\right)=R$
is reached.

\begin{figure}
\begin{centering}
\includegraphics{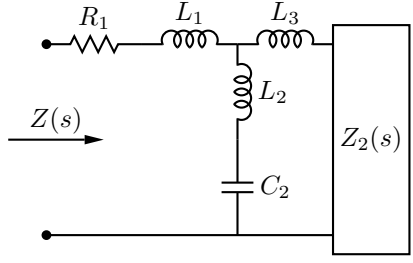} 
\par\end{centering}

\caption{\label{fig:Brune-circuit-extraction}Brune circuit extraction step}
\end{figure}

The circuit in Fig. \ref{fig:Brune-circuit-extraction} potentially
involves negative values for inductances $L_{1}$ and $L_{3}$ \cite{Guillemin}.
However one can replace the T-shaped inductive part of the circuit
in Fig. \ref{fig:Brune-circuit-extraction} with a ``tightly coupled''
inductor as shown in Fig. \ref{fig:Brune-stage-with-ci} where the
inductances are related by

\begin{figure}
\begin{centering}
\includegraphics{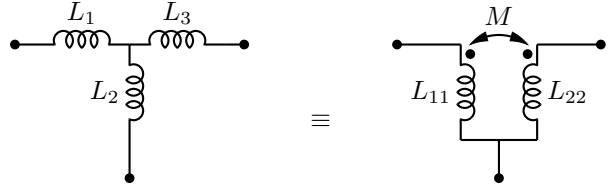} 
\par\end{centering}

\caption{Equivalence of T-shaped inductive circuit in Fig. \ref{fig:Brune-circuit-extraction}
to a coupled inductor\label{fig:Brune-stage-with-ci}}
\end{figure}

\begin{align}
L_{11} & =L_{1}+L_{2}\\
L_{22} & =L_{3}+L_{2}\\
M & =L_{2}
\end{align}
Note that lower terminals of the coupled inductor are short-circuited.
A generic 2-port coupled inductor is shown in Fig. \ref{fig:Generic-2-port-coupled-inductor}
with the following constitutive relations

\begin{figure}
\begin{centering}
\includegraphics[scale=1.25]{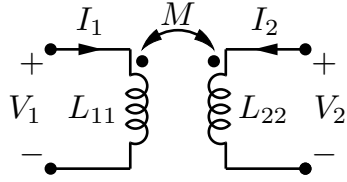} 
\par\end{centering}

\caption{\label{fig:Generic-2-port-coupled-inductor}Generic 2-port coupled
inductor with convention chosen for current directions and voltage
polarities}
\end{figure}

\begin{equation}
\left(\begin{array}{c}
\Phi_{1}\\
\Phi_{2}
\end{array}\right)=\left(\begin{array}{cc}
L_{11} & M\\
M & L_{22}
\end{array}\right)\left(\begin{array}{c}
I_{1}\\
I_{2}
\end{array}\right)
\end{equation}
assuming the conventions shown in Fig. \ref{fig:Generic-2-port-coupled-inductor}
for current directions and voltage polarities. With the current directions
chosen the stored energy in the coupled inductor is given by

\begin{equation}
E=\frac{1}{2}\left(L_{11}I_{1}+2MI_{1}I_{2}+L_{22}I_{2}\right)
\end{equation}

Note that in step (2) above one may find $\omega_{1}=0$ or $\omega_{1}=\infty$
. In case of $\omega_{1}=\infty$ we have the degenerate circuit in
Fig. \ref{fig:degenerate-Brune-stage-C} which corresponds to the
circuit in Fig. \ref{fig:Brune-stage-with-ci} with $L_{1}=L_{2}=L_{3}=0$
. This condition is equivalent to $L'_{k}\rightarrow0$ and $t_{k}\rightarrow0$
. $C_{j}$ in Fig. \ref{fig:degenerate-Brune-stage-C} is given by

\begin{equation}
C_{j}=\underset{s\rightarrow\infty}{\lim}\frac{1}{s\left(Z_{j}-R_{j}\right)}
\end{equation}

\begin{figure}
\begin{centering}
\includegraphics[scale=1.25]{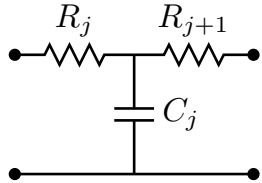} 
\par\end{centering}

\centering{}\caption{\label{fig:degenerate-Brune-stage-C}A degenerate stage in Brune circuit}
\end{figure}

\section{Lossy Foster Method\label{sub:Lossy-Foster-Method}}

Foster's Theorem can be extended to responses with small loss \cite{PMC}.
We start with the partial fraction expansion for $Z\left(s\right)$

\begin{eqnarray}
Z\left(s\right) & = & \underset{k}{\sum}\frac{R_{k}}{s-s_{k}}\label{eq:partial-fraction-expansion-impedance}
\end{eqnarray}
where $R_{k}$'s are residues and $s_{k}$'s are poles. Residues and
poles come in complex conjugate pairs. If we define

\begin{eqnarray}
s_{k} & = & \xi_{k}+j\omega_{k}\\
R_{k} & = & a_{k}+jb_{k}
\end{eqnarray}
Collecting terms corresponding to conjugate pairs

\begin{eqnarray}
Z_{k}\left(s\right)=\frac{R_{k}}{s-s_{k}}+\frac{R_{k}^{*}}{s-s_{k}^{*}} & = & 2\frac{a_{k}s-(a_{k}\xi_{k}+b_{k}\omega_{k})}{s^{2}-2\xi_{k}s+\xi_{k}^{2}+\omega_{k}^{2}}
\end{eqnarray}
One can show that for physical circuits with small loss $\xi_{k}$
and $b_{k}$ are both small quantities \cite{Guillemin}. Hence we
can approximately write

\begin{equation}
Z_{k}\left(s\right)\cong\frac{2a_{k}s}{s^{2}-2\xi_{k}s+\omega_{k}^{2}}\label{eq:lossy-Foster}
\end{equation}
The impedance function of the shunt-resonant circuit as depicted in
Fig. \ref{fig:Shunt-resonant-circuit} is

\begin{figure}
\begin{centering}
\includegraphics[scale=1.25]{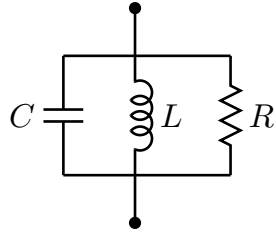} 
\par\end{centering}

\centering{}\caption{\label{fig:Shunt-resonant-circuit}Generic shunt resonant stage in
lossy Foster circuit}
\end{figure}

\begin{equation}
Z\left(s\right)=\frac{\frac{\omega_{0}R}{Q}s}{s^{2}+\frac{\omega_{0}}{Q}s+\omega_{0}^{2}}
\end{equation}
with

\begin{align}
\omega_{0}^{2} & =\frac{1}{LC}\\
Q & =\omega_{0}RC
\end{align}
Hence we see that we can realize the function $Z_{k}\left(s\right)$
in Eq. \eqref{eq:lossy-Foster} by a circuit as in Fig. \ref{fig:Shunt-resonant-circuit}
with

\begin{align}
R & =-a_{k}/\xi_{k}\\
\omega_{0} & =\omega_{k}\\
Q & =-\omega_{k}/2\xi_{k}
\end{align}
and the impedance in Eq. \eqref{eq:partial-fraction-expansion-impedance}
can be realized as in Fig. \ref{fig:Lossy-Foster-Circuit} by a series
connection of stages in Fig. \ref{fig:Shunt-resonant-circuit}.

\begin{figure}[H]
\begin{centering}
\includegraphics[scale=0.8]{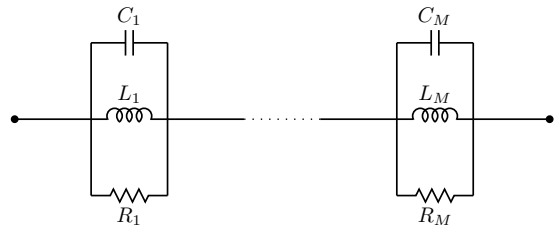} 
\par\end{centering}

\caption{\label{fig:Lossy-Foster-Circuit}Lossy Foster Circuit}
\end{figure}

\end{widetext}
\end{document}